\documentclass[letterpaper,12pt]{article}
\usepackage{authblk}
\usepackage{tabularx} 
\usepackage{amsmath}  
\usepackage{graphicx} 
\usepackage[margin=1in,letterpaper]{geometry} 
\usepackage{cite} 
\usepackage{soul}
\usepackage{hyperref}
\hypersetup{colorlinks,allcolors=black}

\usepackage{blindtext}

\usepackage{lipsum}
\usepackage{fancyhdr}
\begin{document}

\title{AI assisted Malware Analysis: \\ A Course for Next Generation Cybersecurity Workforce}
\author[1]{Maanak Gupta}
\author[2]{Sudip Mittal}
\author[3]{Mahmoud Abdelsalam}
\affil[1]{\small{Dept. of Computer Science,
Tennessee Technological University, Email: mgupta@tntech.edu}}
\affil[2]{Dept. of Computer Science, University of North Carolina Wilmington, Email: mittals@uncw.edu}
\affil[3]{Dept. of Computer Science, Manhattan College, Email: mabdelsalam01@manhattan.edu}

\date{}
\maketitle

\thispagestyle{fancy}%

\begin{abstract}

The use of Artificial Intelligence (AI) and Machine Learning (ML) to solve cybersecurity problems has been gaining traction within industry and academia, in part as a response to widespread malware attacks on critical systems, such as cloud infrastructures, government offices or hospitals, and the vast amounts of data they generate. AI- and ML-assisted cybersecurity offers data-driven automation that could enable security systems to identify and respond to cyber threats in real time. However, there is currently a shortfall of professionals trained in AI and ML for cybersecurity. Here we address the shortfall by developing lab-intensive modules that enable undergraduate and graduate students to gain fundamental and advanced knowledge in applying AI and ML techniques to real-world datasets to learn about Cyber Threat Intelligence (CTI), malware analysis, and classification, among other important topics in cybersecurity. 

Here we describe six self-contained and adaptive modules in ``AI-assisted Malware Analysis.'' Topics include: (1) CTI and malware attack stages, (2) malware knowledge representation and CTI sharing, (3) malware data collection and feature identification, (4) AI-assisted malware detection, (5) malware classification and attribution, and (6) advanced malware research topics and case studies such as adversarial learning and Advanced Persistent Threat (APT) detection. 

\end{abstract}

\section{Introduction}

The use of Artificial Intelligence (AI) and machine learning (ML) to solve cybersecurity problems has been gaining more traction among industry and academia. This data driven automation will enable security systems to identify and respond to cyber threats in real time. The widespread malware attacks on critical issues including cloud infrastructures, government offices or hospitals, together with vast amounts of data generated have necessitated the need for formal education for AI and ML assisted cybersecurity education in universities. The current shortfall of professionals who can use the AI and ML skills for cybersecurity demands development and integration of such curriculum in the computer science and cybersecurity programs.

The need for automation and adaptation has made AI one of the most sought-out skills in the security industry \cite{techgenix}. 
In the year 2019, around 948 government agencies, educational establishments and health-care providers got hit with a barrage of ransomware attacks at a potential cost of \$7.5 billion \cite{crn}. We can anticipate that such attacks on mission critical infrastructure will continue to grow in coming years. AI embedded in analyst augmentation systems will play a crucial role in cybersecurity. 
As mentioned in a recent report \cite{capgemini}, 73\% of large to medium sized organizations are testing use-cases for AI in cybersecurity. Currently, 28\% are using AI embedded security products, while 30\% using proprietary AI algorithms. 42\% currently use, or plan to use AI assisted cybersecurity products. The use of machine and deep learning in cybersecurity is trending upwards, 
with almost two out of three (63\%) organizations planning to employ such products by the end of 2020. The use of AI and Machine Learning (ML) for cybersecurity is one of the most in-demand cybersecurity skill \cite{techgenix}. 

Many universities have a wide variety of courses in AI and cybersecurity, there are still very limited opportunities for students to apply AI in cybersecurity domain. Students need to be educated about the use of AI and ML technologies in security systems so as to identify and respond to threats in real time. 
Researchers have actively developed novel AI and ML solutions for, Cyber Threat Intelligence \cite{mittal2016cybertwitter,mittal2019cyber,khurana2019preventing,piplai2019creating,shakarian2018dark,pingle2019relext,neil2018mining,mittal2017thinking,piplaiknowledge}, Malware Analysis \cite{abdelsalam2018malware,ranade2018using,mcdole2020analyzing, watson2015malware}, Malware Classification \cite{huang2016mtnet}, etc. to prevent and detect cyber-attacks.

Here we describe a course titled ``\textit{AI assisted Malware Analysis}'' to transfer this research to students. Providing students with the knowledge of using AI in malware analysis will be an incredibly powerful tool to bridge the cybersecurity talent gap. It will open up the opportunities for not only cybersecurity focused talent, but also from students across other concentrations like data science or machine learning to apply their skills to solve cybersecurity problems. On the other side, cybersecurity focused students who add AI to their skillset can expect to open many more opportunities in this highly sought-after field.

In the course students are expected to gain fundamental and advanced knowledge in using AI and ML techniques on real-world dataset for cyber threat intelligence, malware analysis, classification among other important topics in the domain.

\section{Course Overview \& Pre-requisites}

\begin{figure}
    \centering
\includegraphics[scale=0.55]{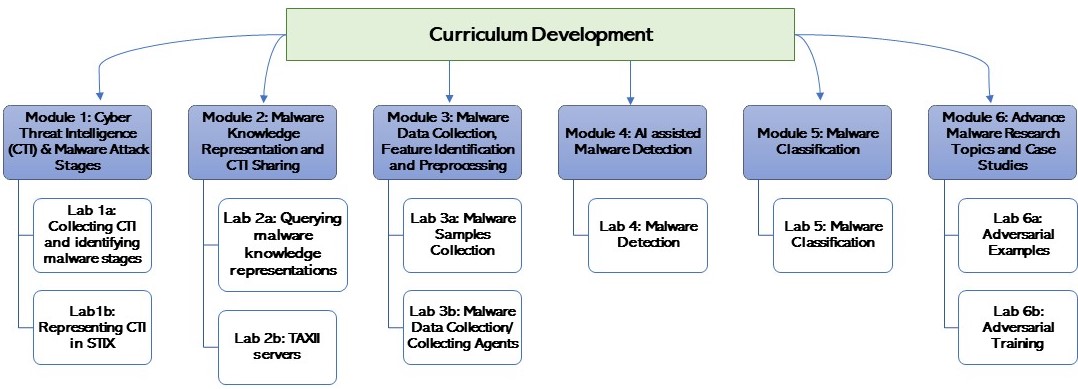}  
\caption{Overview of Course Modules and Labs.}
  \label{fig:module_overview}
\end{figure}

Here we present a summary of the curriculum that will be offered as part of the course: ``Artificial Intelligence assisted Malware Analysis''. 
This course will have 6 modules as shown in Figure \ref{fig:module_overview}, including (1) Cyber Threat Intelligence (CTI) and malware attack stages (2) Malware knowledge representation and CTI sharing (3) Malware data collection and feature identification (4) AI assisted malware detection (5) Malware classification and attribution (6) Advanced malware research topics and case studies. Here each module consists of lectures and lab sessions. Figure \ref{fig:module_overview}, shows a strong bias towards experimentation/lab exercises.

Before enrolling in a course that teaches the proposed modules, students should have completed the following \textbf{pre-requisites} concepts -
\begin{enumerate}
    \item \textit{Introduction to Machine Learning/Artificial Intelligence} - Students are supposed to have taken an undergraduate introduction to machine learning/artificial intelligence, generally taught to computer science majors at various universities. Students are expected to know key concepts and applications of machine learning/artificial intelligence. 
    \item \textit{Introduction to Cybersecurity} - The proposed course expect students to have basic cybersecurity foundations, which are completed in 3 foundational knowledge units (Cybersecurity Foundations,  Cybersecurity Principles, IT Systems Components) as discussed in the  NSA/DHS CAE-CDE designation requirements \cite{caecde}.

\end{enumerate}

\section{Course Module Descriptions}

\subsection{Module 1: Cyber Threat Intelligence (CTI) and Malware Attack Stages}\label{m1}

This module covers topics that describe various stages of a malware attack and how corresponding CTI is created for a particular attack. We first discuss how security analysts describe and differentiate between various malware stages like, reconnaissance, initial compromise, command and control, lateral movement, exfiltration and corruption \cite{TCKC}. Next the students understand how security researchers, to combat these malware based attacks, retrieve malware samples from the `wild'. These samples are then `detonated' in a controlled environment and its behaviour `logged'. Using this behavioral data, security analysts map various malware to known indicators and means of attacks. As a result of such studies, the security analysts produce CTI. We also discuss various sources of CTI like, NIST's National Vulnerability Database (NVD) \cite{NVD}, Common Vulnerabilities and Exposures (CVE) \cite{CVE}, After Action Reports \cite{piplai2019creating}, Social Media \cite{mittal2016cybertwitter}, Blogs and News Sources \cite{satyapanich2020casie}, Dark Web \cite{shakarian2018dark}, VirusTotal \cite{VirusTotal}. 


Students then are able to understand how CTI allows an organization to identify, assess, monitor, and compute a response to cyber threats. CTI includes indicators of compromise; tactics, techniques, and procedures (TTPs) used by threat actors; suggested actions to detect, contain, or prevent attacks; recommended security tool configurations; threat intelligence reports and the findings from the analyses of cyber incidents. Most organizations already produce multiple types of cyber threat information about internal threats as part of their security operations efforts. Students are made to understand that they can first \textit{identify} internally available intelligence and \textit{augment} it by referring to \textit{trusted} external CTI sources. 

Furthermore, we also introduce an existing structured language for CTI, Structured Threat Information eXpression (STIX) \cite{stix2}, developed under the OASIS framework \cite{oasis1}. This is followed by how it can be used for developing malware knowledge representations for AI assisted malware detection techniques (See subsections \ref{m2} and \ref{aim}). 

\begin{itemize}
    \item \textbf{Lab 1a: Collecting CTI and identifying malware stages:} In this lab, students will create a program to collect CTI through various feeds like, NVD JSON 1.1 Vulnerability feed \cite{NVDFEED} and VirusTotal (Public API Endpoint limited to 4 requests per minute) \cite{VirusTotal}. They will then be tasked to identify details about various malware stages in the collected CTI samples. This lab can be done by students in any programming language like python or Java. 
    \item \textbf{Lab 1b: Representing CTI in STIX:} In this lab, participants will implement a system that helps representing CTI in STIX. They will be given multiple CTI samples that need to be represented in STIX. Students will utilize publicly available STIX project \cite{STIXGIT} python libraries to parse, manipulate, generate \cite{STIXPYTHON}, and validate \cite{STIXVALID} STIX content. The lab will teach students to correctly assign intelligence indicators of compromise; tactics, techniques, and procedures (TTPs), etc. to the correct corresponding STIX schema bindings \cite{STIXSCHE}.
\end{itemize}

\subsection {Module 2: Malware Knowledge Representation and CTI Sharing}\label{m2}

Knowledge representation is a field of AI that focuses on designing efficient computer representations that capture information vital for complex problems. In order to create AI assisted malware detection systems, students need to understand various malware representation techniques. In this module, we focus on popular techniques that enable security analysts to store malware data and representations in databases, knowledge graphs, and vector spaces. 

We begin this module by introducing students to a typical malware database schema, based on the STIX schema \cite{STIXSCHE}. We provide students this malware behavioural feature dataset that includes CPU usage, memory usage, disk I/O, network usage, etc. We have access to such a dataset collected as a part of a research project \cite{abdelsalam2018malware}. This also helps us explain to students various malware data collection schemes discussed later in Module 3 (subsection \ref{m3}). 

Students then learn about various AI specific malware representations, which include malware knowledge graphs, ontologies, and vector spaces. These representation techniques have been used extensively to represent behavioral information and CTI, and enable students to understand different knowledge representation techniques for malware from a variety of perspectives \cite{mundie2013ontology, xia2017malware,DING2019101574,pingle2019relext,mittal2016cybertwitter,syed2016uco, mittal2017thinking, DBLP:conf/naacl/RoyPP19, DBLP:conf/semeval/PadiaRSFPPJF18, DBLP:journals/corr/abs-1709-07470}. We also highlight various advantages and disadvantages associated with the use these representation techniques, for example, the use of ontology reasoning mechanisms that are processed over a knowledge graph, experiments have shown that this method has high malicious code detection rate and low false alarm rate \cite{piplai2019creating, DING2019101574}.


We discuss the benefits of cyber threat information sharing \cite{johnson2016guide}. The modules will cover Trusted Automated Exchange of Intelligence Information (TAXII) \cite{TAXII}. TAXII is an exchange framework that allows various organizations to share CTI using the STIX format (See Section \ref{m1}). It enables distribution of CTI among cyber information sharing and analysis organizations. Students are introduced to various CTI sharing models that present disparate use-cases depending on organization policy, like, Hub and Spoke model, Source/Subscriber model, Peer To Peer model \cite{TAXII}. 
\begin{itemize}
    \item \textbf{Lab 2a: Querying malware knowledge representations:} Students will be given access to servers hosting various malware representations like databases, knowledge graphs, and vector models. Students will run specific queries, find similarities and dis-similarities between various malware representations.
    \item \textbf{Lab 2b: TAXII servers:} In this exercise, students will connect to open source TAXII servers \cite{hailataxii,OASISOPEN} and download posted Cyber Threat Intelligence (CTI). Students will also configure a TAXII server themselves \cite{taxiiserver} and post available STIX CTI to their TAXII servers \cite{taxiiclient}. 
\end{itemize}

\subsection{Module 3: Malware Data Collection, Feature Identification and Preprocessing}\label{m3}
In this module, students learn three steps of malware data collection and feature identification: samples gathering, features identification, and data collection. In samples gathering, we discuss ways of acquiring malware samples including honeypots (active and passive) and malware public databases (e.g., VirusTotal \cite{VirusTotal} and VirusShare \cite{virusshare}). In features identification, we discuss commonly used static features like binary n-grams, Control Flow Graphs (CFGs) and static API calls, along with behavioral features like performance metrics, memory information, and system calls. In data collection, we discuss the usage of isolated environments such as sandboxes (e.g., Cuckoo Sandbox) and virtual machines. We will also discuss the limitations of using isolated environments and other alternatives including the use of a live testbed for real world use cases simulation. Further, we discuss host-based and network-based collecting agents as well as virtual machine introspection. The collected dataset with also be used to populate a malware database schema as discussed in Module 2 (subsection \ref{m2}).


The collected data features which are of different scales and categories, require preprocessing. We teach feature normalization and standardization techniques including discussion on categorical data and ways to convert it into numerical data. For features which are in huge numbers, we discuss dimensionality reduction techniques, specifically Principle Component Analysis (PCA) and Independent Component Analysis (ICA).

\begin{itemize}
    \item \textbf{Lab 3a: Malware Samples Collection:} In this lab, students will utilize a public malware database service to acquire working ransomware samples for linux platform. They will be provided an isolated VM where they can ensure that collected samples are working properly.

    \item \textbf{Lab 3b: Malware Data Collection/Collecting Agents:} In the second part, students will write a host-based collecting agent to capture malware behavioral data. This lab will focus on capturing malware system calls. Students will be provided with an isolated VM where they can utilize the samples acquired in Lab 3a to run, test and collect the data using their agent.
\end{itemize}
\subsection{Module 4: AI assisted Malware Detection}\label{aim}
\begin{figure}
  \begin{center}
    \includegraphics[width=\textwidth]{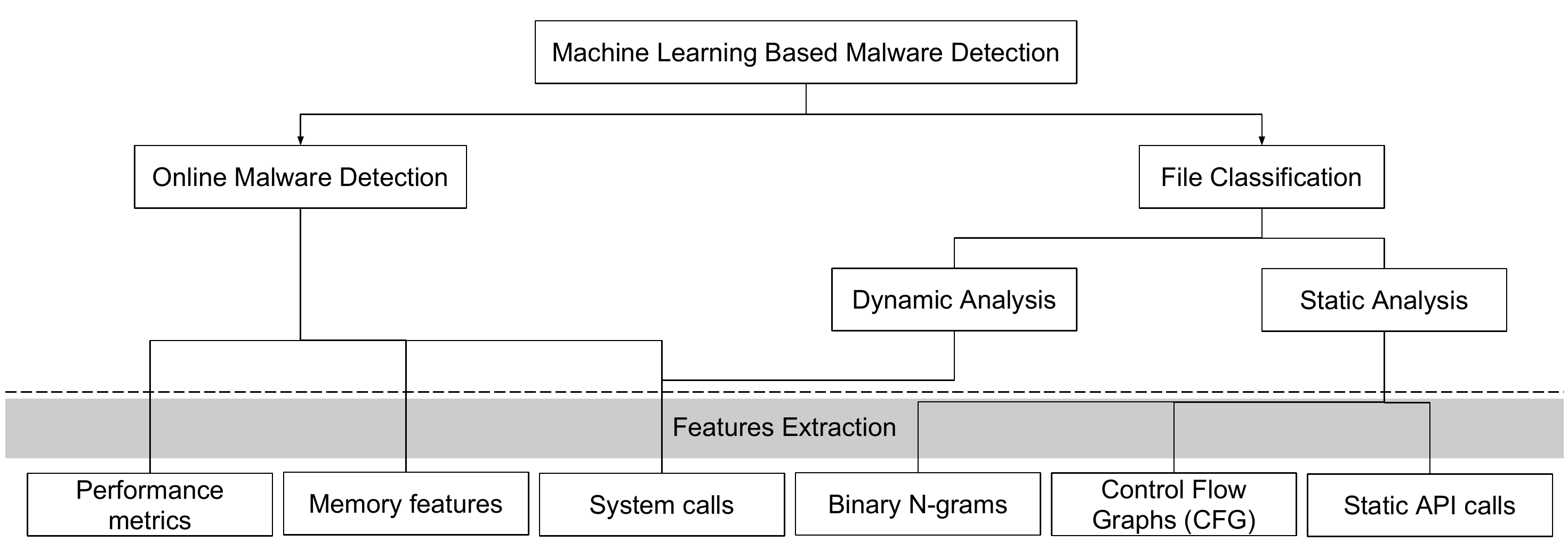}
  \end{center}
  \caption{Overview of AI Assisted Malware Detection}
  \label{fig:malware_detection_overview}
\end{figure}
In this module students learn various AI/ML based malware detection techniques and corresponding extracted features as shown in Figure \ref{fig:malware_detection_overview}. Malware detection techniques can be divided into two categories: online malware detection and file classification. Some of the techniques we discuss in this module include classical algorithms like, support vector machines \cite{watson2015malware}, DBSCAN \cite{chandola2009anomaly}, etc. and state-of-the-art deep neural networks \cite{abdelsalam2018malware}.

We start with file classification techniques including static and dynamic analysis. In static analysis, students learn three major classes of features including: Binary N-grams~\cite{tahan2012mal, abou2004n, kolter2006learning, shabtai2009detection}, Control Flow Graphs (CFGs)~\cite{tobiyama2016malware, kan2018towards, eskandari2012ecfgm} and Static features/Disassembling~\cite{saxe2015deep, seok2016visualized}. Although static analysis techniques are efficient, most recent malware are sophisticated and has polymorphic nature, which hinder the effectiveness of static analysis. To overcome this, students are also taught AI/ML based dynamic analysis techniques, which focus on behavioural aspects of malware. 
To that end, we discuss various tools needed to monitor system processes, filesystem and registry changes and network activity. We provide a use case that focuses on running executables in a controlled environment and observing their behavior, where system/API calls ~\cite{agrawal2018robust, athiwaratkun2017malware, dahl2013large, huang2016mtnet, kirat2015malgene} are mainly used.

File classification approaches can be considered prevention mechanisms, since only executables deemed as benign can be run on the system leaving no chance for malicious executables to run free. However, they can not help detecting malware in an already checked benign application that got infected later on. Henceforth, we will develop modules for online detection approaches, which will help understand and develop ability to continuously monitor the entire system for the presence of malicious activities. We will include different features used in this approach which are more dynamic and time dependent such as performance metrics~\cite{demme2013feasibility, abdelsalam2017clustering, abdelsalam2018malware, abdelsalam2019online}, memory features~\cite{ozsoy2015malware, xu2017malware} or run-time system/API calls~\cite{dawson2018phase, luckett2016neural, alazab2011zero}.


\textbf{Lab 4: Malware Detection: }
In this lab, students will be tasked to use the previously collected data in Lab 3b (subsection \ref{m3}) and select one AI/ML algorithm to apply on this dataset. They will apply data pre-processing step as necessary depending on the chosen AI/ML algorithm. Students will be provided with a GPU-enabled machines to help speeding up the AI/ML model training. Additionally, students will report their findings with rationalization of their choices, supporting graphs and proper explanation.
\subsection {Module 5: AI assisted Malware Classification}
Malware is used for many purposes such as: stealing data, asking for ransoms, creating backdoors, tracking user activity, and spreading spam, to say the least. Because of the ubiquitous nature of malware, it is important not only to detect the presence of malware but also the type (e.g., worm, trojan, backdoor, rootkit, ransomware, etc.) and the function of that malware. This can help in faster remedy actions. As such, students learn the different malware types and how to apply different machine learning algorithms to classify malware into known families.

Classification is not an easy task, since it is not as simple as placing a malware in an unambiguous class. In most cases, real-world malware has a wide variety of nefarious capabilities and propagation mechanisms. This makes classification harder and unpredictable. Even when families of malware share similarities, minor modifications in a malware can cause confusion to the classifier. Malware classification goes beyond simply classifying the malware based on its name. In this module, different methods are discussed including classic classification, deep learning and clustering based approaches. Students also learn how to choose the right classification scheme depending on specific use cases. This can be achieved using NIST's~\cite{souppaya2013guide} outlined steps an organization can take to develop a malware classification scheme to prioritize specific incidents.

\textbf{Lab 5: Malware Classification: } In this lab, students train a model to classify malware into 9 families based on the Kaggle dataset \cite{kagglemalware, ronen2018microsoft}. They will be required to employ AI/ML algorithms and a subset of features extracted from the dataset which they believe will increase the accuracy of the classifier, providing their technical reasoning behind such choices. Additionally, students will be given a VM with a fixed defined specification to run their experiments, which will limit their choices of the AI/ML algorithm architecture as well as the number of features to use, given the lab deadline. Students will utilize python scikit-learn \cite{scikit-learn}, PyTorch \cite{NEURIPS2019_9015} and/or Tensorflow \cite{tensorflow2015-whitepaper} framework for developing AI/ML algorithms. 

\subsection{Module 6: Advance Malware Research Topics and Case Studies}
In this module, advance and evolving topics related to AI assisted cybersecurity are discussed. 
First, students learn about adversarial machine learning attacks, where models can be fooled through malicious input. We discuss how adversaries can poison the model to make wrong classifications, different kinds of adversarial ML attacks, including the black box and white box attacks. Data Poisoning and evasion techniques are also discussed. Defense strategies to combat adversarial ML attacks will be covered. These include primarily two types of defense strategies: 1) reactive: detect adversarial examples after deep neural networks are built; 2) proactive: make deep neural networks more robust before adversaries generate adversarial examples. We discuss three reactive countermeasures (Adversarial Detecting, Input Reconstruction, and Network Verification) and three proactive countermeasures (Network Distillation, Adversarial (Re)training, and Classifier Robustifying).

Another important topic to be discussed is Advance Persistent Threats (APTs). We discuss APT progression including the network infiltration, expansion and extraction. With around 73\% of the organizations rate their detection capability as inadequate, it is important to focus on detection techniques. We discuss AI based APT detection and analysis techniques. Fuzzing has been used traditionally to find software bugs by randomly feeding data into a target program until one permutation reveals a vulnerability. Applying AI and ML models to fuzzing enables it to become more efficient and effective. We discuss such techniques and relevant tools. 

In this module, students are also introduced to case studies discussing recent malware attacks and applications of AI for malware analysis.
This includes case studies about recent ransomware attacks against cities and critical establishments across the U.S. It highlights how such attacks were orchestrated and how AI/ML assisted detection mechanisms discussed in the earlier modules can be used to restrict their effects. 
\begin{itemize}
    \item \textbf{Lab 6a: Adversarial Examples}: In this lab, students are tasked with generating adversarial examples that bypass the detection techniques they built in the previous labs (\ref{m3} and \ref{aim}) by utilizing cleaverhans \cite{cleverhanslib} library. This will help them in learning how such examples are crafted and used in practice.
    \item \textbf{Lab 6b: Adversarial Training}: In this lab, students are expected to apply adversarial training, one of the easiest and most effective approaches to defend against adversarial attacks. Adversarial training is an approach where a model is retrained using adversarial examples so that the model will not be fooled by such examples.
\end{itemize}




\section{Conclusion}
In this article, we provide an outline of a cybersecurity course, which will discuss state of the art skill sets with respect to the use of AI and ML for malware analysis. We firmly believe that such a course when introduced at various universities and educational institutions will produce graduates and future workforce who will be well versed and equipped to prevent, detect and mitigate against sophisticated cyberattacks.

\section*{Acknowledgement}

This work was supported by National Science Foundation awards 2025682, 2025685, and 2025686.

\bibliographystyle{unsrt}
\bibliography{references}

\end{document}